# HEAVY AND SUPERHEAVY ELEMENTS PRODUCTION IN HIGH INTENSIVE NEUTRON FLUXES OF EXPLOSIVE PROCESS[*]

**Yu. S. Lutostansky**[†]

*National Research Center "Kurchatov Institute", Moscow, 123098 Russia*

[†]*E-mail: lutostansky@yandex.ru*

**V. I. Lyashuk**

*Institute for Nuclear Research, Russian Academy of Science, Moscow, 117312 Russia;*

*National Research Center "Kurchatov Institute", Moscow, 123098 Russia*

**I. V. Panov**

*Institute Institute for Theoretical and Experimental Physics, Moscow, 117259 Russia*

*National Research Center "Kurchatov Institute", Moscow, 123098 Russia*

**Abstract:**
Mathematical model of heavy and superheavy nuclei production in intensive pulsed neutron fluxes of explosive process is developed. The pulse character of the process allows dividing it in time into two stages: very short rapid process of multiple neutron captures with higher temperature and very intensive neutron fluxes, and relatively slower process with lower temperature and neutron fluxes. The model was also extended for calculation of the transuranium yields in nuclear explosions takes into account the adiabatic character of the process, the probabilities of delayed fission, and the emission of delayed neutrons. Also the binary starting target isotopes compositions were included. Calculations of heavy transuranium and transfermium nuclei production were made for "Mike", "Par" and "Barbel" experiments, performed in USA. It is shown that the production of transfermium neutron-rich nuclei and superheavy elements with $A \sim 295$ is only possible in case of binary mixture of starting isotopes with the significant addition of heavy components, such as long-lived isotopes of curium, or californium.

## 1. Introduction

The understanding of heavy elements formation process in nature is based on the progress in nuclear physics and astrophysical scenario as well. Observation of abundances for elements gives us the final results of different nucleosynthesis processes in different objects during billions of years. That is why additional experimental information can give very important knowledge of nucleosyntheis details. This information can be received from the analysis of nucleosynthesis in explosions [1] and simulation of such processes.

[*] This work is supported by grants 12-02-00955, 13-02-12106 ofi-m, 14-22-03040 ofi-m of the Russian Foundation for Basic Research and Swiss National Science Foundation grant no IZ73Z0_152485 SCOPES.

The production of transuranium elements in nature takes place in powerful neutron flux owing to reactions of multiple neutrons capturing and the following β-decays. This process of rapid nucleosynthesis (the *r*-process) is realized, for example, in the explosions of supernova stars, where neutron density exceeds $10^{20}$ neutron/cm$^3$ at temperatures of ~ $10^9$ K. Pulsed nature of the process of nucleosynthesis under conditions of supernovae explosions and lasting a few seconds is inseparably linked with time-dependent external conditions – neutron flux and temperature. The model of transuranium isotopes production under stationary conditions of nucleosynthesis was developed in 1985 [2] – 1990 [3,4].

In artificial conditions, the *r*-process is realized in nuclear explosions, which produce neutron fluencies above $10^{24}$ neutron/cm$^2$ in a time of ~ $10^{-6}$ s. This short-time process may be called the "*prompt*" pulse *r*-process or *pr*-process. Investigations of transuranium nuclei production in such a process were performed in 1952 – 1964 years in the USA in thermonuclear tests. In these "experimental" thermonuclear explosions the neutron flux on the $^{238}$U target was from $(1.2 – 1.8) \cdot 10^{24}$ ("Mike") [5] to ~ $6.6 \cdot 10^{24}$ ("Par", "Barbel") neutron/cm$^2$ [6, 7]. Transuranium isotopes (up to $^{255}$Fm) were registered for the first time in the "Mike" thermonuclear explosion in 1952 [5]. At that time, studies were performed in the United States to examine the possibility of transuranium elements synthesizing under the conditions of nuclear explosion (the "*Plowshare*" program). The most complete data on transuranium yields up to $A = 257$ were obtained in the "Par" experiment [7]. In order to increase yields of transuranium isotopes and to search for isotopes with mass numbers $A > 257$, it were conducted the experiments that produced high neutron fluencies. In the "Hutch" test, a maximum fluence of $2.4 \times 10^{25}$ neutrons/cm$^2$ was achieved [7], but no isotopes with $A > 257$ were found. Nuclides created during nucleosynthesis pulse are very neutron-rich and fast decaying. Analysis of the nuclides composition, however, is completed about 2 days after the pulse and in this time, nuclei with high numbers of neutrons undergo multiple beta- and α-decay [2, 7]. In this work we analyze the possibility of the transfermium neutron-rich nuclei production in pulsed neutron fluxes of explosive process including and the thermonuclear explosions.

Modeling the *r*-process under astrophysical conditions [2], we must consider the (*n*, γ)-reaction of neutron radiation capture and the inverse (γ, *n*)-process; induced and spontaneous nuclear fissions; the β-decay of neutron-rich nuclei accompanied by processes of delayed one and two neutrons emission (β, *n*), (β, 2*n*) [8], delayed fission (β, *f*) [9], α-decay [10] and so on. To perform calculations, we must establish neutron and neutrino fluxes, temperature conditions, and the parameters of more than 4000 nuclei. In the experimental pulse of *pr*-process, the model description of the synthesis of heavy elements allows important simplifications: processes of radiation neutron capture and β-decay are strongly separated in time [$(t_{(n,\gamma)} \approx 10^{-6}$ s$) << t_\beta$], which can significantly limit the range of nuclei involved in the *pr*-process. Also it is possible the simplification by neglecting the neutrino interaction processes. Processes accompanying β-decay of neutron-rich nuclei can be inserted in the model on the second stage of calculation.

## 2. Dynamic nucleosynthesis model

In the dynamic model of heavy elements nucleosynthesis in nature (see, for example, [11]), the concentration of nuclide $N_A(Z,N,t)$ depends on time *t*, as well as the neutron fluxes $N_n(t)$, gamma rays $N_\gamma(t)$ and neutrinos $N_\nu(t)$. Since the change in concentration of $N_A(Z,N,t)$ decays make contributions from neighboring nuclei with $A \pm 1$, and (*n*, γ), (γ, *n*) and (ν, γ) reaction in neighboring nuclei, to determine the $N_A(Z,N,t)$ it is necessary to solve a system of more than 4000 equations, depending on the time. The solution of this system of equations is an independent complex mathematical task that requires large computer resources. The task is complicated by the fact that, in these equations, the velocities of all nuclear reactions are time dependent.

As was shown by numerous calculations, the concentration of nuclides are highly dependent on the used nuclear data, the number of which more than tens of thousands. And the majority of these data we must predict, because the involved in the process of neutron-rich short-lived nuclei in the most have not been studied and have not even been observed experimentally. The dependence of the results of calculations the concentrations of producing in the *r*-process nuclides from the nuclear data examined in detail in [12, 13, 14] by the calculations of heavy and superheavy elements formation.

Dynamic *r*-process can be represented as a two-stage process, both as passing and the temperature process. In the first short stage, the initial temperature exceeds $10^9$ K, and this is "hot" *r*-process [14]. In the second, longer stage, the so-called "cold" *r*-process realized [16] with temperature $T \ll 10^9$ K. It was found that if the ambient temperature is significantly below 1 MeV ($T^9 \sim 0.1$), then the *r*-process may be more effectively for the superheavy elements production since the *r*-process path is shifted closer to the border of the neutron nuclear stability and the nucleosynthesis wave moves faster [14]. This fact is interesting for "*prompt*" *pr*-process, the conditions of which probably correspond to the "cold" *r*-process than "hot".

An important aspect of the theory is checking the reliability of predictions of nuclear data. For the *r*-process the measured abundance characteristics of transuranium elements in the nuclear explosions [1, 7] are practically the only source of experimental data (except for the observations in the matter of the solar system and the stellar atmospheres). We have carried out within the framework of a large mathematical computer code for the astrophysical nucleosynthesis calculation showed that the observed in the explosions the curve of heavy nuclei abundance is well described, when irradiated with neutrons matter consists of a mixture of 2 - 3 chemical elements.

Later in this paper we consider a special model that allows to verify the calculations used in astrophysical nuclear data on experimental data pulse *pr*-process. In this paper we used modern cross sections and nuclear reaction rates obtained for the simulation of the formation of heavy and superheavy elements in astrophysical conditions [12]. Half-life periods, probabilities of emission for one and two delayed neutrons (DN), probabilities of delayed fission (DF) for neutron-rich isotopes were calculated taking into account the resonance structure of β-strength function, which obtained from the finite-Fermi system theory [15 – 17].

## 3. Binary adiabatic model for *pr*-process

During modeling the *r*-process under artificial conditions [18], i.e. *pr*-process, for nuclear (thermonuclear) explosions, the above mentioned significant simplification were made because the processes of neutron capture and β-decay are separated in time. This model is used for calculation of transfermium elements production. As starting isotopes on the first stage the binary composition of $^{238}$U and $^{239}$Pu were used. For the calculations of the transfermium elements production we used also the (Pu + Cm) and (Pu + Cf) compositions of initial target nuclei. The *pr*-process can be classified as "cold" two-step process with a temperature $T \ll 10^9$ K where the time "tail" is determined by the flow of DN and is many times smaller than the main stream hard neutrons [19].

The process of production the transuranium elements under the conditions of nuclear (thermonuclear) explosions can be described by the system of equations for concentrations for the nuclide concentration $N_z^n(t)$ depending on time [3, 18] as well as neutron flux $F(E, t)$. The rates of (*n*, γ), (*n*, 2*n*), (*n*, 3*n*), and (*n*, *f*) were taken into account as well as β- and α-decays and spontaneous fission.

We can further simplify the model taking into account the features of the process over time and the contribution from reactions. The chain reaction proceeds for ~$10^{-7}$ s, and the duration time of multiple radiation captures of neutrons does not exceed $10^{-6}$ s [19]. The contribution from β- and α-decays and spontaneous fission are negligibly small, since their rates are much smaller than the rate of *n*-capture.

The scheme for transuranium isotope creation in intense explosive neutron fluxes of the *pr*-process is shown in Fig. 1, as compared to the slower (*s*-process) trunsuranium production that occurs, e.g., in nuclear reactors.

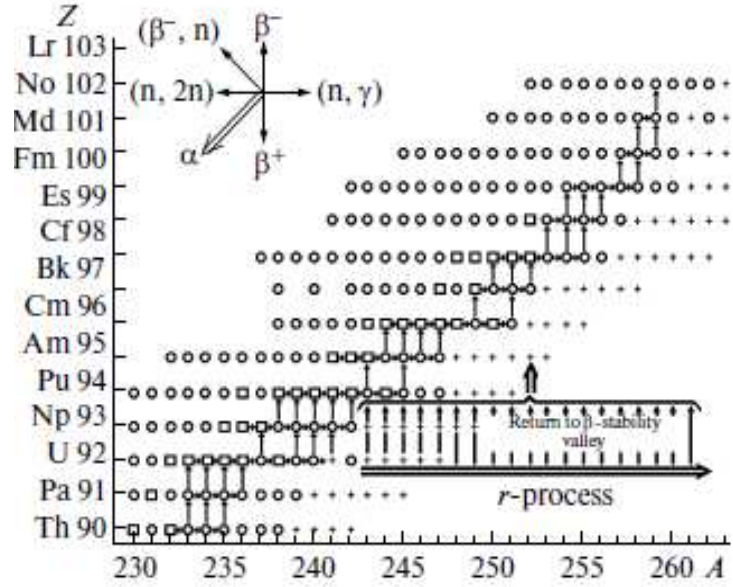

**Fig. 1.** Scheme of actinide creation in artificial slow (*s*-process) and pulse *r*-process –prompt rapid process (*pr*-process) of nucleosynthesis: □ – nuclei with $T_{1/2} \geq 1$ year; O – nuclei with $T_{1/2} < 1$ year; + – predicted neutron-rich nuclei from the NDS JAEA data base; the line denotes the path of the *s* -process at a flux density of neutrons ~ $(10^{14} – 10^{16})$ neutron /(cm$^2$ s).

Further development involves introducing elements of dynamics to this static model, which accounts for the change in radiation capture rates over the time (~$10^{-6}$ s) of multiple neutron capture conduction. After the ending of fission chain reactions (~$10^{-7}$ s), the matter scatters rapidly [20] and in the time range of the *pr*-process (~$10^{-6}$ s) the increase of the volume of highly heated plasma leads to fast cooling of the matter involved in the motion.

The dependence of temperature on time for the time interval $[t_A – t_B]$ is determined by the range $(T_i – T_{i+1})$ of decrease in temperature, and it is assumed that the radial expansion velocity of a heated substance is constant in the interval $[t_A – t_B]$, and the dependence of temperature on volume is adiabatic [18, 20]: $T = (const/V)^{\gamma - 1}$. And in our calculations, it was assumed that the adiabatic index – $\gamma$ ranged from 1.5 to 2.0 [21]; a reduction in temperature was fixed in the range from 60 to 1 keV.

## 4. Losing factor

Following pulse prompt nucleosynthesis, neutron-rich isotopes undergo β-decay, upon which two main processes leading to a change in concentration are possible:

(β, n)-delayed emission of neutrons, and (β, f)-delayed fission. These processes lose isotopes in isobaric chains with the constant mass number $A$ and, as a result, the distribution of the isotope yields according to the mass number $A$ changes considerably to the end of the *pr*-process. The losing effect summarized by the isobar chain gives a relative reduction in concentrations for a given $A$ and is expressed as the $L(A)$ coefficient (the losing factor, where $L(A) \leq 1$) and the concentration of isotopes with given $A$, calculated at the moment of the end of multiple captures, must be multiplied by the factor $R(A) = 1 - L(A)$. Generally speaking, the $L$-factor value depends on time because the current concentrations of the decaying neutron-rich nuclei are time-dependent. In our case, the time interval from the beginning of the *pr*-process and up to the experimental analysis of the final concentrations is too large in comparison with the $T_{1/2}$ of neutron-rich nuclei. And so we ignore the time dependence of $L$-factor.

Figure 2 shows the resulting coefficients of the drop in concentrations $L(A)$ and contributions from the (β, n) and (β, f) processes to $L(A)$ coefficients.

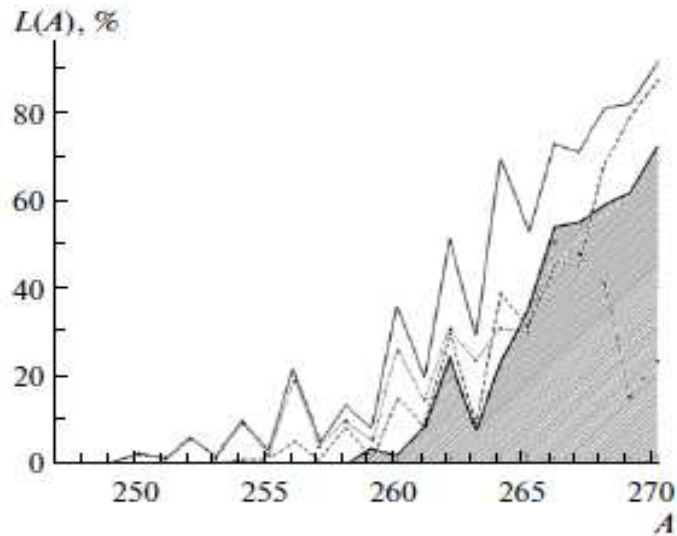

**Fig. 2.** Losing factor $L(A)$ (%) in isobaric chains for initial isotopes of U and Pu. For uranium, the dotted line is the contribution from delayed fission; the dashed line is the contribution from delayed neutron emission; and the solid line is the summed effect. For plutonium, the bold line (shaded portion) is the cumulative effect.
.

In calculations with the initial isotope $^{238}$U, the losing factor $L_U(A)$ increases at $A = 252, 254, 256, 258$ i.e. at even $A$ isotopes. In this case, the main contribution comes from the (β, *f*) processes on even *A* neptunium isotopes, which explains the observed odd-even abnormality in the yields distribution. Spontaneous fission and α-decay were not considered in our calculations of *L*(*A*), but it is significant for mass number $A \geq 256$ of isotopes $^{256}$Cf, $^{258}$Fm, and heavier nuclei [22]. For the initial isotope $^{239}$Pu, the $L_{Pu}(A)$ factor is substantial for $A > 260$ and is systematically smaller than the $L_U(A)$ factor, since uranium and neptunium isotopes are excluded from the calculations. Data on the binding energy of neutrons and fission barriers from [23, 24] were used in our calculations of the *L*(*A*) coefficients.

In the case of a plutonium target, the losing factor $L_{Pu}(A)$ is very small for neutron-rich isotopes with $A < 260$. The $L_U(A)$ and $L_{Pu}(A)$ factors for the initial irradiated uranium and plutonium isotopes differ strongly (see Fig.2), and this is explained by the different amounts of nuclides participating in the formation of the losing factor.

Much smaller losing factor will be for curium or californium targets. In this case the main losing effect will be caused by of spontaneous fission and α-decay of the neutron-rich nuclei. And the interval of nuclei masses included in *pr*-process up to $A = 295$ will be smaller than for uranium target, so the probability of transfermium elements production will be higher.

## 5. Even-odd anomaly

The transuranium isotopes yields calculations are shown in Fig. 3, relative to the experimental yields measured in the "Par" test, where all yields up to $A = 257$ were obtained for the first time [7]. The results were normalized using nuclide yield with $A = 245$, as in [7]. The calculated data up to $A = 270$ were obtained in an integral neutron flux of ~ $6 \times 10^{24}$ neutron/cm$^2$ for pure uranium $^{238}$U(100%) and uranium–plutonium targets with the initial concentration $^{238}$U(95%) + $^{239}$Pu(5%). In the experimental yields with $A > 250$, the effect of even–odd inversion was observed as a break at $A \approx 250$ in the characteristic saw-toothed yields, and as inversion of the yields at $A > 250$ (Fig. 3).

To explain the even-odd effect in the model, the influence of the delayed processes – DF and DN is considered as a correction due to the *L*(*A*) losing factor, which increases for the even *A*-nuclides as *A* increases (see Fig. 2) and acts in the right direction, approaching the calculated results for experimental concentrations registered after explosion. The model was further complicated by the simultaneous inclusion of two isotopes, $^{238}$U and $^{239}$Pu [18], in the composition of the initial target. It should be noted that the dependences of the calculated yields on *A* for $^{238}$U and $^{239}$Pu are in opposite phases (as are the respective cross sections of (*n*, γ)-reactions), which improves agreement with the experimental data in the region of the inversion effect at *A* > 250. The resulting yields calculations for "Par" experiment are presented in Fig. 4 (the horizontal dashed line at the level of unity corresponds to complete agreement with the experimental data) as relations to the experimental data for the initial targets of $^{238}$U(100%) and $^{238}$U(95%) + $^{239}$Pu(5%), and losing factors $L_U(A)$ and $L_{Pu}(A)$.

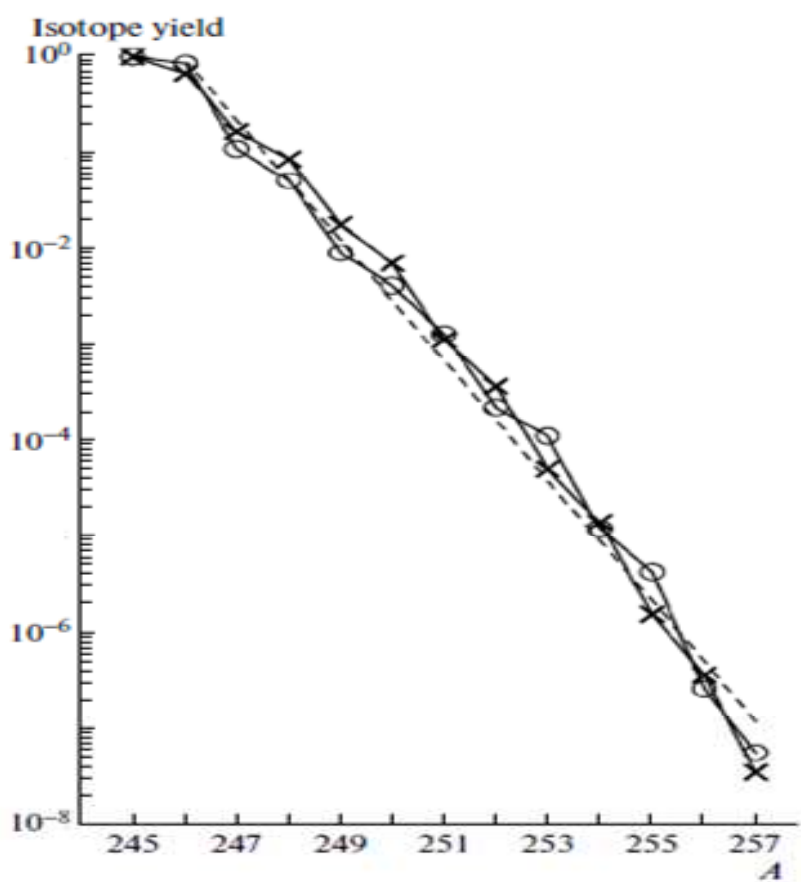

**Fig. 3.** Isotope yields in the "Par" experiment:
○ – experimental data;
× – calculated results without consideration of the process in time dynamics;
the dashed line denotes the fitting of the calculated isotope yields by the function:
$Y = \exp(-1.442A + 354.56)$.

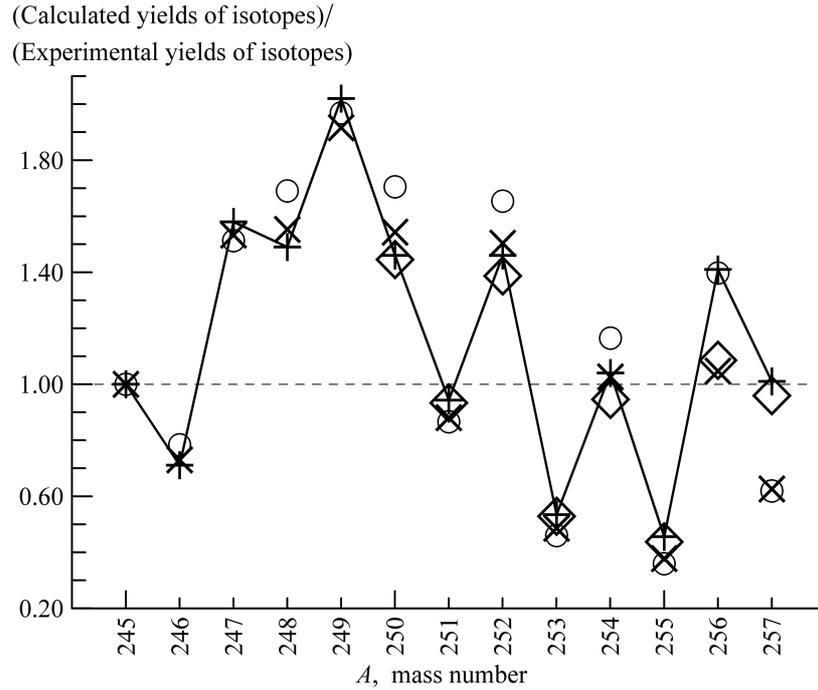

**Fig. 4**. Ratio of calculated to experimental values of isotope yields in the "Par" test for mass numbers $A$ = 245–257: ○- calculated results for the model with $^{238}$U in the target; × - calculated results with allowance for the losing factor for the target with $^{238}$U as a single isotope; + - calculated data (solid line) for the model with the binary target ($^{238}$U, 95%; $^{239}$Pu, 5%) in which losing factor is ignored; ◊ - calculated results for the model with the binary target and with the losing factor.

So, we can explain the even-odd anomaly mainly by influence of beta-delayed fission and in smaller part, by influence of plutonium impurity in starting isotopes. This is a good test for the further calculations of transfermium elements production.

## 6. On the possibility of the formation of transfermium elements by pulse of neutrons

As it was shown earlier [25], the possibility of the formation of transfermium elements in *pr*-process using $^{238}$U or (U + Pu) targets is very small. Including heavier chemical elements in a target during pulse neutron nucleosynthesis could be a promising way of synthesizing transfermium neutron-rich nuclei and superheavy (SH) elements. Yields of heavy nuclides were therefore evaluated for the inclusion of small additives of $^{248}$Cm and $^{251}$Cf isotopes (available during reactor operation) into the initial uranium target. Here the losing factor $L(A)$ was calculated without the inclusion of spontaneous fission and α-decay. It should be noted that the $L(A)$ factor in this case is smaller than for the above mentioned uranium and plutonium targets, since the number of neutron-rich nuclides taking part in β-decay is smaller and they are located not far from the region of beta-stability. Therefore the beta-delayed processes for these neutron-rich nuclei are not so intensive compared with the very short-living nuclides producing from the uranium target.

The inclusion of small masses of curium into the seed mixture with $^{238}$U (Fig. 5) enables us to increase the yields of isotopes with mass numbers $A > 250$ by an one and even two orders of magnitude (for $A = 254$ at a 5% concentration of $^{248}$Cm). Further the yields caused by $^{248}$Cm become approximately equal to yield from U-part at: $A = 256\text{-}257$ for Cm – 0.1%; $A = 259\text{-}260$ for Cm – 0.5%; $A = 265\text{-}266$ for Cm – 5.0%. For larger $A$-number the yields from Cm are decrease and curves *4, 5* and *6* lay close to $^{238}$U (100%) target yield.

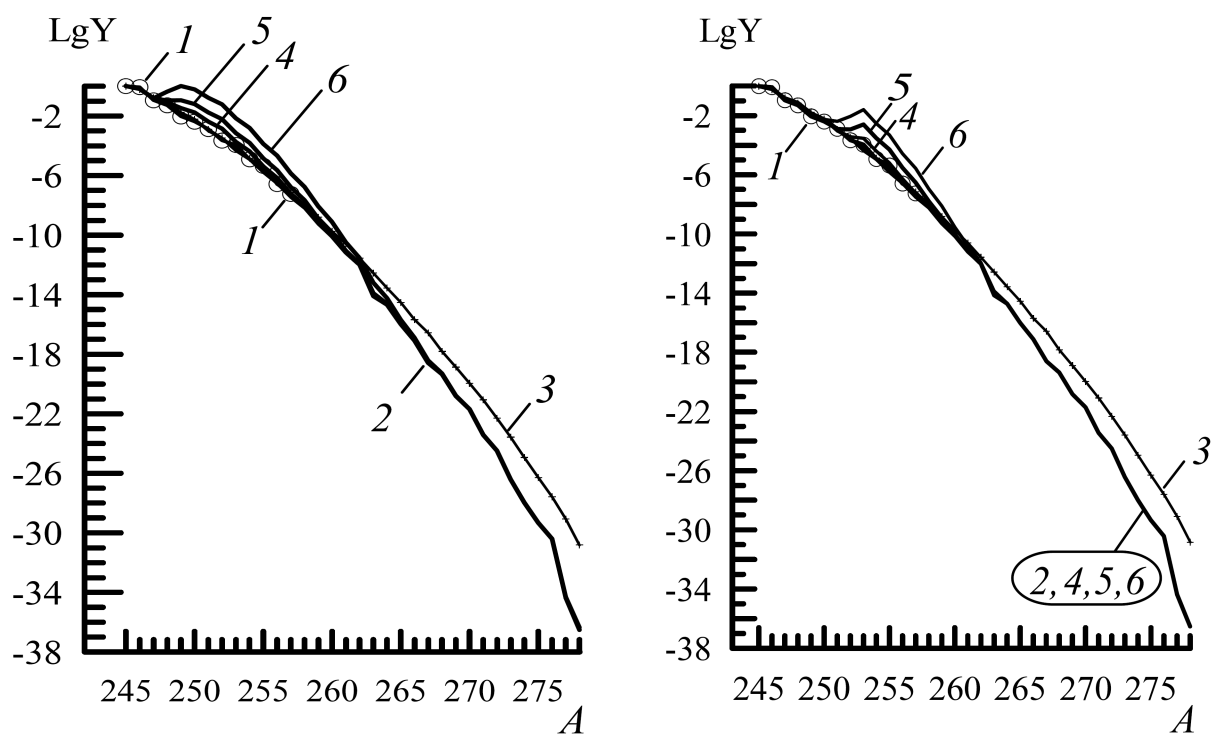

**Fig. 5**. Comparison of the calculated yields $Y(A)$ for (U + Cm) target on the left figure, (U + Cf) target on the right figure, U and (U + Pu) targets at a flux of 6x10$^{24}$ neutrons/cm$^2$. Curves: *1* – ○ - "Par" exp. data; *2* – $^{238}$U(100%); *3* – $^{238}$U(95%) + $^{239}$Pu(5%). For (U + Cm) target: curves *4, 5, 6* for concent-ration of Cm – 0.1%, 0.5% and 5%, respectively. For (U + Cf) target: curves *4, 5, 6* for concentration of Cf – 0.001%, 0.01% and 0.1%, respectively.

At a concentration of 0.001% $^{251}$Cf there is thus a more than fivefold increase in the yield of isotopes with the mass number $A = 253$. For larger concentration – 0.1%

$^{251}$Cf the yield of $A=253$ isotopes increases more than five hundreds. At $A = 256, 257$ for Cf – 0.001%, $A = 259, 260$ for Cf – 0.01% and $A = 262–264$ for Cf – 0.1% the isotope yields from californium became close to yield from uranium-part in the target. And for further increase $A$-number the total yields from (U + Cf) target approach to the uranium yield.

We can conclude that introducing small doses of $^{248}$Cm and especially of $^{251}$Cf into the $^{238}$U mixture strongly increase the yield at wide $A$-interval. But as can be seen from Fig. 5 using Cm isotope additives at small concentrations in a target mainly consisting of uranium does not have a great effect on transuranium nuclides yields as does a (U + Pu) binary target. And the same is for Cf isotope [25]. In order to enhance the effect, the concentration of additives should be far higher, but this does not make much sense in experiments with a destructible target. So, the better for SH nuclei production may be the (Pu + Cf) target.

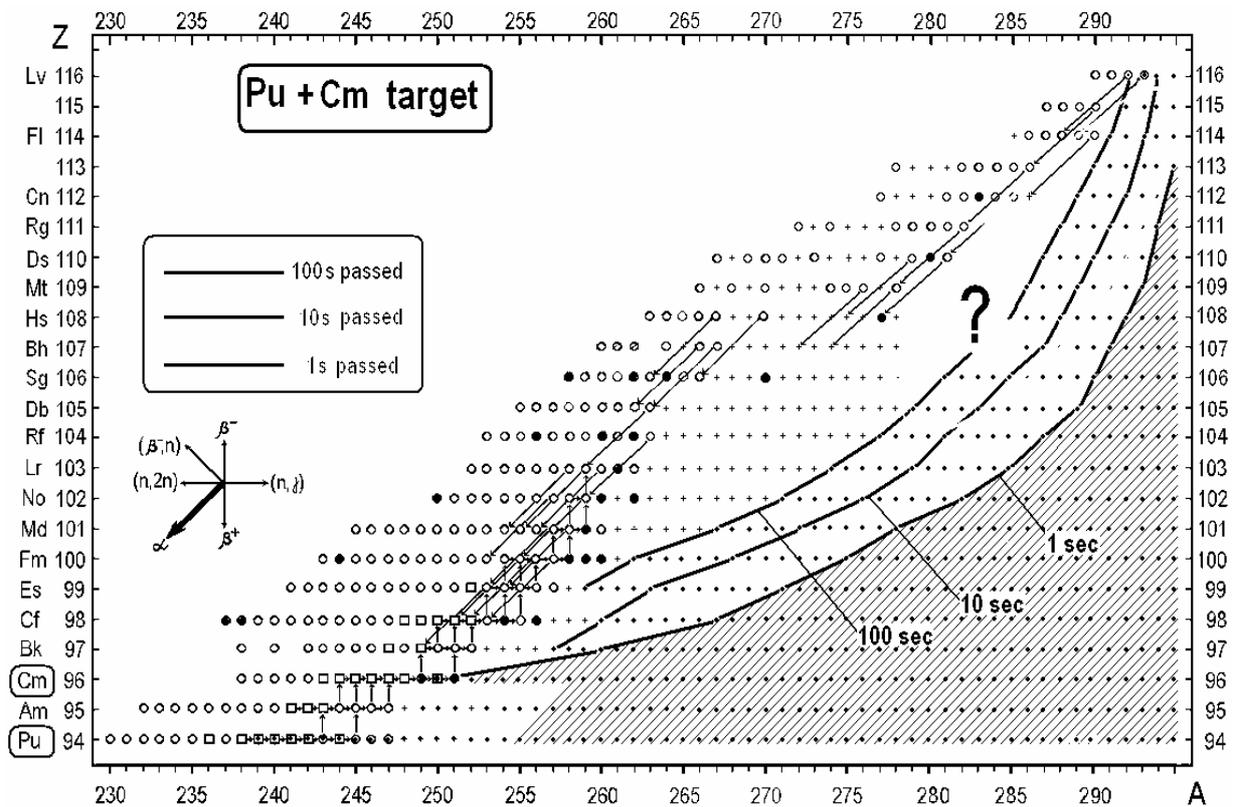

**Fig. 6**. Distribution for producing of neutron-rich nuclei for (Pu + Cm) target during one second after neutron pulse, 10 sec and 100 sec after neutron pulse.

Figure 6 shows the distribution of producing neutron-rich nuclei for (Pu + Cm)

target during one second: after neutron pulse, 10 sec and 100 sec after neutron pulse. In these calculations we use time-dependent losing factor $L(A, t)$. The analogous calculations were performed for the (U + Pu) target. In case of (U + Pu) target, the ultra short-lived neutron-rich nuclei with $A = 280 – 295$ decay rapidly and in the pause time after the neutron pulse $t_p > 100$ sec the nuclei only with $Z < 100$ will remain. So, we can conclude that transfermium isotopes will not be produced.

However, when we use (Pu + Cm) target (Fig. 6), the produced neutron-rich nuclei decay much slowly since they have smaller beta-decay energy and already in the pause time after the neutron pulse $t_p > 100$ sec remain nuclei with $Z > 100$, that are much more promising for the transfermium isotopes production. The further fate of radioactive transmutations we did not considered since it is necessary to use the theory of spontaneous fission [22, 26] and α-decay [10] that will make calculations more complicated.

## 7. Conclusion

For the calculations of transuranium isotopes yields in pulse neutron fluxes the adiabatic binary model of transuranium elements production is developed. The calculations of the transuranium isotopes yields up to $A = 295$ in pulsed neutron fluxes of high intensity were performed with start isotopes: $^{238}$U, $^{239}$Pu, $^{248}$Cm and $^{251}$Cf.

Comparison of yields calculations up to $A = 257$ for binary targets $^{238}$U + $^{239}$Pu for different proportions, with experimental data were carried out to the "Par", "Barbel" and "Mike" USA thermonuclear explosions. The obtained results are in good agreement with the experimental data.

Experimental data of transuranium yields in nuclear explosions revealed anomalous odd-even effect, which manifests itself in the mass number $A > 250$. This "losing" effect is explained by the delayed fission process in $^{252,254,256}$Np isotopes, formed after radiation of uranium by intense neutron flux. It is shown that the losing effects $L(A)$-factor increases greatly for $A > 260$ nuclei associated with the U – Np – Pu neutron-rich isotopes. Calculations with this effect lead to the better agreement with experimental data of "Par" event. The agreement of the calculated isotopes yields with the experimental data is up to 50%.

It is shown that nuclei with $A \approx 270$ can be obtained in the "Par" experiments with the yields $\sim 10^{-22}$ using uranium target, and – with the yields $\sim 10^{-20}$ using binary (U + Pu) targets. Use of (U + Cm) and (U + Cf) target can significantly (up to two orders of magnitude) increase the yields of isotopes in the wide range of $A \sim (250 - 260)$. For more heavier nuclei with $A \approx 278$ the yields can be evaluated as extremely small values as $10^{-31}$ for (U + Pu) target and $10^{-37}$ for U target. Such low concentrations are impossible to detect by modern experimental methods. Moreover, these nuclides decay rapidly.

Superheavy elements production with $A \sim 290$ is possible only by using the original binary mixture of isotopes with a noticeable addition of heavy components, such as long-lived isotopes curium or californium. It was found that transfermium neutron-rich nuclei with $A = 280–295$ decay faster and in the pause time after the neutron pulse $t_p > 100$ s remain only nuclei with $Z < 100$ in the case of (U + Pu) target. But in the case of (Pu + Cm) or (Pu + Cf) target with the pause time $t_p > 100$ sec after neutron pulse the nuclei with $Z > 100$ will remain, that is much more promising for the transfermium isotopes production.

**Acknowledgments**

The authors are grateful to E.E. Sapershtein, A. Sobiczewski, V.N. Tikhonov and S.V. Tolokonnikov for their assistance and helpful discussions.

The work was partially supported by the Russian Foundation for Basic Research Grants no. 12-02-00955, 13-02-12106 ofi-m (part 6), 14-22-03040 (part I) ofi-m and Swiss National Science Foundation grant no IZ73Z0_152485 SCOPES (part 7).